\documentclass[twocolumn,secnumarabic,amssymb, nobibnotes,prd,superscriptaddress]{revtex4}

\usepackage[utf8]{inputenc}
\usepackage{graphics,graphicx}
\usepackage{amsmath,amssymb,amsfonts,latexsym,cancel}
\usepackage{multirow}

\usepackage{bm}

\begin{document}

\title{Tidal deformability and radial oscillations of anisotropic polytropic spheres}

\author{Jos\'e D. V. Arba\~nil}
\email{jose.arbanil@upn.pe}
\affiliation{Facultad de Ciencias F\'isicas, Universidad Nacional Mayor de San Marcos, Avenida Venezuela s/n Cercado de Lima, 15081 Lima,  Peru}
\affiliation{Departamento de Ciencias, Universidad Privada del Norte, Avenida el Sol 461 San Juan de Lurigancho, 15434 Lima,  Peru}
\author{Grigoris Panotopoulos}
\email{grigorios.panotopoulos@ufrontera.cl}
\affiliation{Centro de Astrof\'isica e Gravita\c{c}\~ao-CENTRA, Departamento de F\'isica, Instituto Superior T\'ecnico-IST,
Universidade de Lisboa-UL, Avenida Rovisco Pais, 1049-001 Lisboa, Portugal}
\affiliation{Departamento de Ciencias F{\'i}sicas, Universidad de la Frontera, Casilla 54-D, 4811186 Temuco, Chile}

\date{\today}

\begin{abstract}
We compute the equilibrium, the fundamental eigenfrequency of oscillations modes, and quadrupolar tidal deformability of anisotropic polytropic spheres. These studies are respectively performed through the numerical solution of the Tolman-Oppenheimer-Volkoff equation, Chandrasekhar radial oscillation equations, and nonlinear first-order Riccati equation for tidal deformability, all modified from their original version to include the anisotropic effects. For the polytropic exponent $\gamma=2$ and the anisotropic model of Cattoen, Faber, and Visser, we show that the anisotropy could be reflected in the radial pressure, energy density, speed of sound, radial stability, and tidal deformability.
\end{abstract}

\maketitle


\section{\label{sec:level1} Introduction}

Compact objects \cite{textbook,review1,review2}, such as for instance neutron stars, white dwarfs or quark stars, which arise during the final stages of stellar evolution, are unique probes of properties of matter under exceptionally extreme conditions, which cannot be reproduced on earth-based experiments. As they are the denser objects in the Universe (after black holes), they comprise excellent cosmic laboratories to study and constrain non-standard physics and modified theories of gravity. Matter in the interior of those objects is characterized by ultra-high densities, for which the usual description of stellar plasmas in terms of Newtonian fluids is inadequate. Therefore, very dense compact objects are relativistic in nature, and as such they are properly described only within Einstein’s General Relativity (GR).

\smallskip

When studying relativistic astrophysical objects the authors usually focus on stars made of isotropic fluids, where the radial pressure $p_r$ equals the tangential pressure $p_t$. Celestial bodies, however, are not always made of isotropic matter. As a matter of fact, under certain conditions the fluid can indeed become anisotropic. The review article by Ruderman \cite{paper1} considered for the first time that possibility: In that work the author made the observation that in a very dense medium anisotropies may arise due to relativistic particle interactions. The investigation of the impact of anisotropies on the properties of relativistic stars received a boost by the subsequent work of \cite{paper2}. Indeed, anisotropies may arise in many different contexts involving dense matter media, such as phase transitions \cite{paper3}, pion condensation \cite{paper4}, or in the presence of type 3A super-fluid \cite{paper5}. See also \cite{Ref_Extra_1,Ref_Extra_2,Ref_Extra_3} for more recent works on the topic, and references therein. In those works relativistic models of
anisotropic strange quark stars were studied, while the energy conditions were found to be satisfied. In particular, in \cite{Ref_Extra_1} an exact analytic solution was obtained, in \cite{Ref_Extra_2} an attempt was made to find a singularity free solution to Einstein’s field equations, and in \cite{Ref_Extra_3} the Homotopy Perturbation Method was employed, which is a tool to tackle Einstein’s field equations. Besides, alternative approaches have been introduced capable of inducing anisotropies onto known isotropic seed solutions \cite{Ovalle:2017fgl,Ovalle:2017wqi}.

\smallskip

The inspiral and subsequent relativistic collision of two objects in a binary system, and the gravitational wave signal emitted during the whole process, contain a wealth of information regarding the nature of the colliding bodies. The imprint of the equation-of-state within the signals emitted during binary coalescences is mainly determined by adiabatic tidal interactions, characterized by a set of coefficients, known as the tidal Love numbers and the corresponding deformability. 

\smallskip

The theory of tidal deformability was introduced in Newtonian gravity first by Love \cite{Love1,Love2} more than a century ago, with the purpose of understanding the yielding of the Earth to disturbing forces. In the case of a spherical body, Love introduced two dimensionless numbers to describe the tidal response of the Earth. To be more precise, the first number, $h$, describes the relative deformation of the body in the longitudinal direction (with respect to the perturbation), while the second number, $k$, describes the relative deformation of the gravitational potential.

\smallskip

The consideration of self-gravitating compact objects requires a relativistic theory of tidal deformability, which was developed in \cite{flanagan,hinderer,damour,Lattimer,poisson} for spherically symmetric neutron stars and black holes. Naturally, the key deformability parameter is the relativistic generalisation of $k$, since the role of the gravitational potential is now played by the metric tensor. Then, tensor perturbations of a spherically symmetric geometry fall into two classes: even parity (or electric/polar) and odd parity (or magnetic/axial). Hence, there are electric and magnetic relativistic tidal Love numbers, $k^E$ and $k^B$, respectively. Moreover, each one may be decomposed in spherical harmonics of index $l$, with $l$ being the angular degree, thus introducing $k_l^E$ and $k_l^B$.

\smallskip

The latest cosmological data \cite{planck} indicate that non-relativistic matter in the Universe is dominated by dark matter. Despite the success of the concordance cosmological model at large scales ($\sim 1 Mpc$) based on the cold dark matter paradigm, a series of shortcomings at galactic and sub-galactic scales (i.e. $\sim kpc$ or less) persists, such as the core/cusp problem and the missing satellite problem - for excellent reviews see e.g. \cite{review3,review4}. Those problems may be tackled if dark matter consists of ultralight scalar particles with a mass $m \ll eV$ \cite{Hui:2016ltb}. In those models, ultralight bosons can cluster forming macroscopic Bose-Einstein condensates with the mass of the Sun or even larger. Those self-gravitating clumps for spinless bosons are called scalar boson stars \cite{BS1,BS2,BS3,BS4,Harko,Mielke2,BS5,BS6,BS7,Liebling,BS8}. Boson stars are intrinsically anisotropic, and under certain circumstances they can be described by a polytropic equation-of-state.

\smallskip

Asteroseismology has been a powerful tool to infer the inner structure of the Sun and other similar stars, allowing astronomers to have a detail characterization of the micro-physics and fluid dynamical processes taking place in their interiors, for instance, nuclear reactions, equation-of-state, differential rotation rate and meridional circulation \cite{pulsating1a,pulsating1aa}. Moreover, those techniques are now being extended to the study of the inner structure of compact objects \cite{pulsating1c,pulsating1d,pulsating1e}. 
Those new methods of diagnostic provide us with a robust way to search for hints of non-conventional physics inside stars, such as the existence of dark matter \cite{pulsating1b,pulsating2c} or alternative theories of gravity \cite{pulsating2a,pulsating2b}.
Therefore, in this work by computing the oscillations of those new class of stars, i.e., the frequencies and eigenfunctions of the radial oscillations of pulsating stars, e.g. \cite{stellarpul1,stellarpul2}, we can learn about their composition as well as the equation-of-state of the strongly interacting matter, since the precise values of the frequency modes are very sensitive to the underlying physics, composition and inner structure of the star, see e.g. \cite{vath_chanmugam1992, pulsating2, pulsating3, pulsating4, pulsating5, lugones2010, flores2017, pulsating8, sagun_2020} and references therein.

\smallskip

In the present work we compute the tidal Lover numbers $k_2^E$ as well as the radial oscillation frequencies for non-rotating polytropic spheres made of anisotropic matter having boson stars in mind. The plan of our work is the following: In the next section we briefly review the structure equations describing hydrostatic equilibrium of relativistic stars as well as the equations for the perturbations of pulsating objects, and we summarize the theory regarding tidal Love numbers. In the third section we discuss the equation-of-state and the factor of anisotropy considered here, while in Section IV
we display and discuss our main numerical results regarding several different properties of the polytropic spheres. Finally, we close our work with some concluding remarks in section V. We adopt the mostly positive metric signature, $(-,+,+,+)$, and we work in geometrized units, where the speed of light in vacuum as well as Newton's constant are set to unity, $c=1=G$.

\section{Stellar equilibrium equations and radial pulsation equations}\label{section2}

\subsection{Stellar equilibrium equations}\label{stellareq}

The anisotropic matter making up the compact object is described by a stress-energy tensor as follows:
\begin{equation}
T_\nu^\mu = {\rm Diag}[-\rho, p_r, p_t, p_t],
\end{equation}
where $\rho$ represents the energy density, while $p_r$ and $p_t$ being the radial and tangential pressure of the fluid, respectively. Equivalently, instead of $p_t$ we may work with the anisotropic factor, $\sigma$, defined by
\begin{equation}
\sigma= p_t - p_r.
\end{equation}

The background line element to seek spherically symmetric static objects, in Schwarzschild-like coordinates $(t,r,\theta,\phi)$, is given by
\begin{equation}\label{line_element}
ds^2 = -e^{\nu} dt^2 +e^{\lambda} dr^2 + r^2 (d \mathrm{\theta^2} + \mathrm{sin^2 \theta \: d \phi^2}),
\end{equation}
where the functions $\nu=\nu(r)$ and $\lambda=\lambda(r)$ depend on the radial coordinate only.

The hydrostatic equilibrium equations used to analyze relativistic stars within GR are obtained by using the Einstein field equation and the line element \eqref{line_element}. This set of equations, also known as the Tolman-Oppenheimer-Volkoff (TOV) equations \cite{paper2} (see also \cite{tolman,oppievolkoff}), are:
\begin{eqnarray}
\hspace{-0.5cm}&&m' =4 \pi r^2 \rho,\label{eq_mass} \\
\hspace{-0.5cm}&&p_r' =- (\rho + p_r)\left(\frac{m+4 \pi r^3 p_r}{r^2 e^{-\lambda}}\right)+ \frac{2 \sigma}{r}, \label{eq_p}\\
\hspace{-0.5cm}&&\nu'=2\left(\frac{m+4 \pi r^3 p_r}{r^2 e^{-\lambda}}\right),\label{eq_g00} 
\end{eqnarray}
with 
\begin{equation}
e^{-\lambda}=1-\frac{2m}{r}.
\end{equation}
The primes $(\,'\,)$ stand for derivation with respect to $r$ and $m(r)$ represents the mass inside a radius $r$.

Equations \eqref{eq_mass}-\eqref{eq_g00} are integrated imposing at the centre of the star ($r=0$) the initial conditions:
\begin{eqnarray}
m(0)=0,\quad \rho(0)=\rho_c,\quad p_r(0)=p_c,\nonumber\\
\lambda(0)=0,\quad\nu(0)=\nu_c,\quad {\rm and}\quad \sigma(0)=0,
\end{eqnarray}
with $\rho_c$ and $p_c$ being the central energy density and the central radial pressure, respectively. Since at the surface of the star ($r=R$), the interior solution is connecting with the exterior Schwarzschild vacuum solution, the following condition must be satisfied 
\begin{equation}
p_r(R) = 0.
\end{equation}
At this point, the interior and the exterior metric are related as
\begin{equation}
e^{\nu(R)}=e^{-\lambda(R)}=1 - \frac{2M}{R},
\end{equation}
where $M$ and $R$ are the mass and the radius of the star, respectively.

\subsection{Radial pulsation equations}

The radial stability equations are obtained by perturbing both the space-time and the fluid variables, preserving the spherical symmetry of the background object. The perturbations are introduced in both the field equations and the stress-momentum tensor conservation equations, by maintaining only the first-order terms (see \cite{dev_gleiser_2003,gleiser_dev_2004}). This equation can be placed into two first-order equations, one equation for the relative radial displacement $\xi$ and other one for Lagrangian perturbation $\Delta p_r$ \cite{arbanil_malheiro_2016}. Hence, the radial stability of a compact star is determined through the following set first-order equations: 
\begin{eqnarray}
\hspace{-0.6cm}&&\xi'=\frac{\xi}{2}\nu'-\frac{1}{r}\left[\frac{2\xi\sigma}{\Gamma p_r}\left[\frac{\Gamma p_r}{p_r+\rho}+1\right]+3\xi+\frac{\Delta p_r}{\Gamma p_r}\right],\label{eq_ro_1}\\
\hspace{-0.6cm}&&\Delta p_r'=e^{\lambda-\nu}(p_r+\rho)\omega^2\xi r-4\xi p_r'+\frac{8\sigma\xi}{r}+\frac{2\delta\sigma}{r}\nonumber\\
\hspace{-0.6cm}&&-\left(\frac{\nu'}{2}+4\pi re^{\lambda}(p_r+\rho)\right)\Delta p_r+\frac{\xi r(p_r+\rho)}{4}\nu'^2\nonumber\\
\hspace{-0.6cm}&&-8\pi(p_r+\rho)r\xi e^{\lambda}(\sigma+p_r),\label{eq_ro_2}
\end{eqnarray}
with $\Gamma=\left(1+\frac{\rho}{p_r}\right)\frac{dp_r}{d\rho}$ and $\omega$ being the adiabatic index and the eigenfrequency, respectively.

Eqs. \eqref{eq_ro_1} and \eqref{eq_ro_2} are integrated from along the radial coordinate. To take regular solution in the center of the object, it is considered that for $r\rightarrow0$, the second term of the right-hand side of Eq. \eqref{eq_ro_1} must vanish. Thus: 
\begin{equation}
(\Delta p_r)_{\rm center}=\left[-2\xi\sigma\left[\frac{\Gamma p_r}{p_r+\rho}+1\right]-3\xi\Gamma p_r\right]_{\rm center}.
\end{equation}
At this point, for normalized eigenfunctions, we take $\xi(r=0)=1$. In turn, at the star's surface, where radial pressure is zero, $p_r(R)=0$, the next condition is required:
\begin{equation}\label{delta_p_s}
(\Delta p_r)_{\rm surface}=0.
\end{equation}

\subsection{Tidal deformability}

The theory of tidal Love numbers may be found for instance in \cite{hinderer,damour,Lattimer} for isotropic stars, and \cite{biswas2019} for anisotropic stars. By definition, the tidal deformability parameter $\lambda$ and the dimensionless tidal deformability $\Lambda$ are related to the tidal Love number $k^E_2$ as follows
\begin{eqnarray}
&&\lambda \equiv \frac{2}{3} k^E_2 R^5,\\
&&\Lambda \equiv \frac{\lambda}{C^5},
\end{eqnarray}
with $C=M/R$ representing the compactness of the star.

In terms of $C$, the tidal Love number is computed to be \cite{hinderer,Lattimer}
\begin{eqnarray}
&&k^E_2 = \frac{8C^5}{5} \: \frac{K_{o}}{3  \:K_{o} \: \ln(1-2C) + P_5(C)}, 
\label{elove}\\
&&K_{o}=(1-2C)^2 \: [2 C (y_R-1)-y_R+2],\\
&&y_R \equiv y(r=R),
\end{eqnarray}
where $P_5(C)$ is a fifth order polynomial given by
\begin{eqnarray}
&&P_5(C) = 2 C \: [4 C^4 (y_R+1) + 2 C^3 (3 y_R-2) + \nonumber\\
&&2 C^2 (13-11 y_R) + 3 C (5 y_R-8) -3 y_R + 6]. 
\end{eqnarray}
The function $y(r)$ satisfies a Riccati differential equation, 
\begin{equation}
ry' = -y^2 + [1-r P] y - r^2 Q,
\end{equation}
and the functions $P=P(r)$ and $Q=Q(r)$ are computed in terms of the background quantities as follows
\begin{eqnarray}
\hspace{-1.0cm}&&P = \frac{2}{r} + e^{\lambda} \left[ \frac{2 m}{r} + 4 \pi r (p_r - \rho) \right],\\
\hspace{-1.0cm}&&Q = 4 \pi e^{\lambda} \left[ 4 \rho + 8 p_r + \frac{p_r + \rho}{A c_s^2} (1+c_s^2)   \right] -\nu'^2 -6 \frac{e^\lambda}{r^2},
\end{eqnarray}
where $A = 1+\frac{d\sigma}{dp_r}$, and $c_s^2 = \frac{dp_r}{d\rho}$ is the radial speed of sound, see e.g. \cite{biswas2019} for more details.

\section{Equation-of-state and anisotropic profile}

In this section we motivate the expressions of the equation-of state (EoS) as well as of the form of the factor of anisotropy considered in the present work, following closely \cite{PM-GP}. 

Boson stars are self-gravitating configurations made of either spin-zero fields, called scalar boson stars \cite{Mielke2} or spin-one fields (vector bosons), called Proca stars \cite{Proca1,Proca2}. The highest mass of scalar boson stars neglecting self-interactions was obtained in \cite{BS1,BS2}, and afterwards in \cite{BS3,BS4} it was demonstrated that self-interactions may significantly affect the mass.

A canonical complex scalar field, $\Phi$, is described by an action corresponding to the Einstein-Klein-Gordon system \cite{Liebling}
\begin{eqnarray}
S & = & \int d^4 x \sqrt{-g} \left( \frac{R}{16 \pi} + \mathcal{L}_M  \right), \\
\mathcal{L}_M & = & - g^{\mu\nu} \partial_\mu \Phi \partial_\nu \Phi^* - V(|\Phi|),
\end{eqnarray}
where $g$ is the determinant of the metric tensor $g_{\mu \nu}$, $R$ is the corresponding Ricci scalar, $\mathcal{L}_M$ is the Lagrangian corresponding to the matter content, and $V$ is the self-interaction scalar potential.

For static, spherically symmetric configurations, i.e. if the star does not possess rotation speed, we make for the scalar field the ansatz \cite{Liebling}
\begin{equation}
\Phi(r,t) = \phi(r)\exp({-i \varpi t}),
\end{equation}
with $\varpi$ being a real parameter identified with the oscillation frequency.

It is worth noticing that although the scalar field itself depends on time, the corresponding stress-energy tensor is time independent. Therefore, Einstein's field equations take the usual form for an anisotropic fluid, for which the energy density is computed to be \cite{Cardoso1,Cardoso2}
\begin{equation}
\rho = \varpi^2 e^{-\nu} \phi^2 + e^{-\lambda} \phi'^2 + V(\phi),
\end{equation}
while the radial and tangential pressures are found to be 
\begin{eqnarray}
p_r & = & \varpi^2 e^{-\nu} \phi^2 + e^{-\lambda} \phi'^2 - V(\phi),\label{p_r_def}\\
p_t & = & \varpi^2 e^{-\nu} \phi^2 - e^{-\lambda} \phi'^2 - V(\phi). \label{p_t_def}
\end{eqnarray}

Notice that the radial pressure is different than the tangential one, and therefore a boson star is always anisotropic. The factor of anisotropy is defined by
\begin{equation}
\sigma \equiv p_t - p_r = - 2 e^{-\lambda} \phi'^2 < 0.
\end{equation}
Moreover, it turns out that in the case of boson stars the factor of anisotropy is always negative.

Under certain conditions, however, the anisotropy may be ignored, and so in that case the system may be viewed as an object made of isotropic matter. A concrete model in which the scalar potential is assumed to be
\begin{equation}
V(|\Phi|) = m^2 |\Phi|^2 + \frac{\lambda}{2} |\Phi|^4,
\end{equation}
with $m$ being the mass of the scalar field, and $\lambda$ being the self-interaction coupling constant, was studied e.g. in \cite{maselli/2017}, where the authors considered the following EoS \cite{BS3}:
\begin{equation}
p_r = \frac{\rho_0}{3} \left( \sqrt{1+\frac{\rho}{\rho_0}}  - 1 \right)^2,
\end{equation}
with $\rho_0$ being a constant given by
\begin{equation}
\rho_0 = \frac{m^4}{3 \lambda}.
\end{equation}
This EoS describes boson stars that are approximately isotropic, provided that the condition
\begin{equation}
\frac{\lambda}{4 \pi} \gg m^2,
\end{equation}
holds \cite{maselli/2017}. 

It is worth-mentioning that the well-known results are recovered in the two extreme limits, namely
\begin{equation}
p_r \approx  \frac{\rho^2}{12 \rho_0}, \ \quad \rho \ll \rho_0,
\end{equation}
for dilute stars \cite{ChavHarko}, and 
\begin{equation}
p_r \approx \frac{\rho}{3}, \ \quad \rho \gg \rho_0,
\end{equation}
in the ultra relativistic limit.

Quite generically, in a dilute and cold boson gas the details of the self-interaction potential do not matter, as long as the system is characterized by a repulsive, short range self-interaction. The equation-of-state of a Bose-Einstein condensate is found to be of the polytropic form
\begin{equation}
p_r=K \,\rho^{2},
\end{equation}
corresponding to a polytropic exponent $\gamma=2$, while the constant of proportionality, $K$, is given in terms of the mass of the particles, $m$, and the scattering length, $a$, as follows \cite{DarkStar1,DarkStar2}
\begin{equation}
K = \frac{2 \pi a}{m^3}.
\end{equation}

In the following for convenience we set $K=\frac{z}{B}$ setting $z=0.01$ and $B=66 \, [\rm MeV/fm^3]$.

Finally, following \cite{cattoen2005,horvat2011,silva2015,folomeev2015}, we shall consider for the factor of anisotropy the following expression
\begin{equation}\label{sigma}
\sigma = \kappa p_r(1-e^{-\lambda}),
\end{equation}
where $\kappa$ is called a dimensionless prefactor. This expression ensures that the anisotropy has the correct dimensions and vanishes both at the center and at the surface of the star. Furthermore, since the anisotropic profile contains variables of the fluid and space-time, in the non-relativistic regime, where $1-e^{-\lambda}<<1$, it is expected that $\sigma\sim0$. Since radial pressure is greater than tangential, see Eqs. \eqref{p_r_def} and \eqref{p_t_def}, the anisotropy factor is always negative. Therefore, in order to have $\sigma<0$, we consider $\kappa<0$ due to the functions $p_r$ and $1-e^{-\lambda} $ on the right side of Eq. \eqref{sigma} are always positive. Thus, in the following numerical analysis we will consider $\kappa$ in the range between $-2$ and $0$.

\section{Numerical results}


\begin{figure*}
\centering
\includegraphics[width=8.5cm]{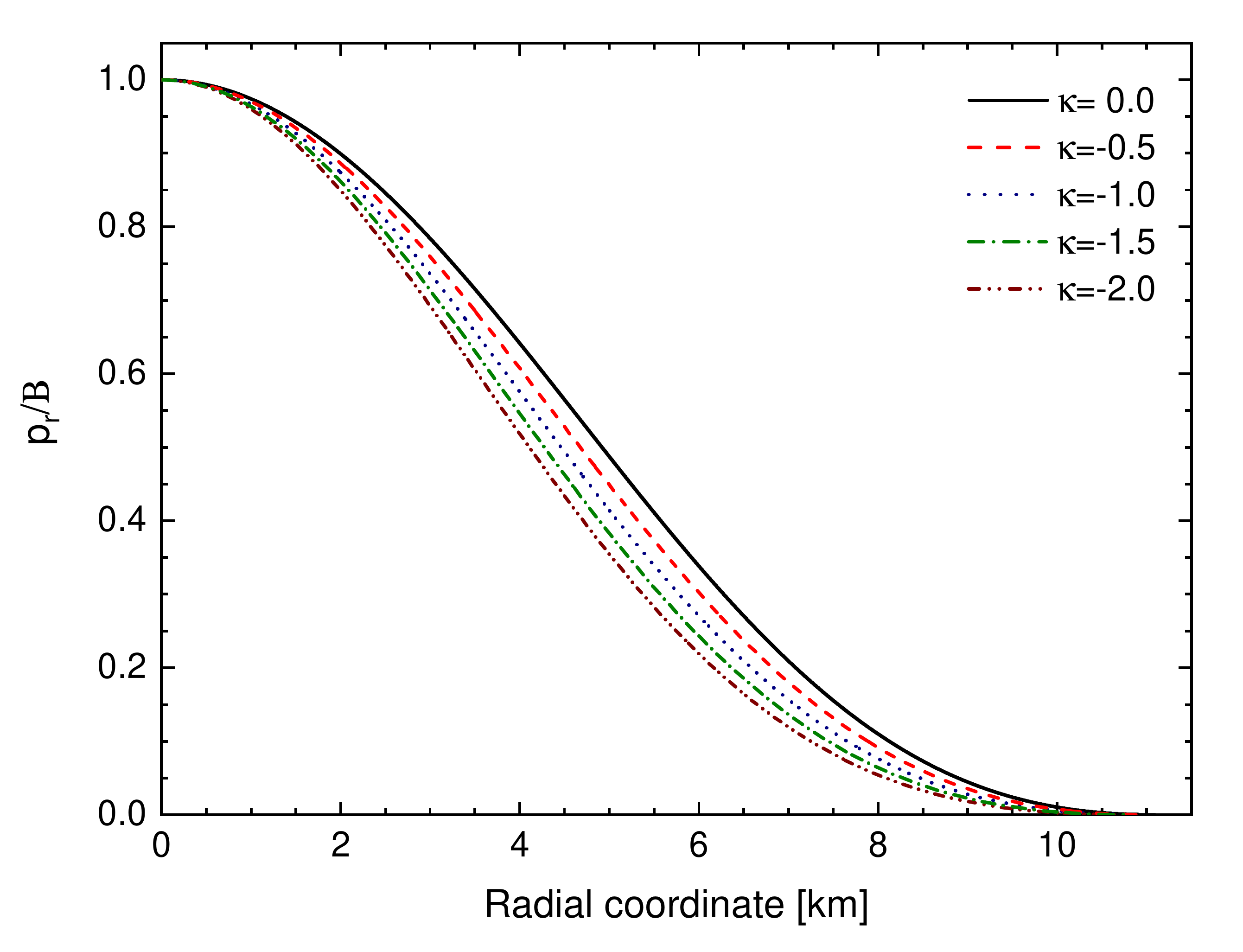} 
\includegraphics[width=8.5cm]{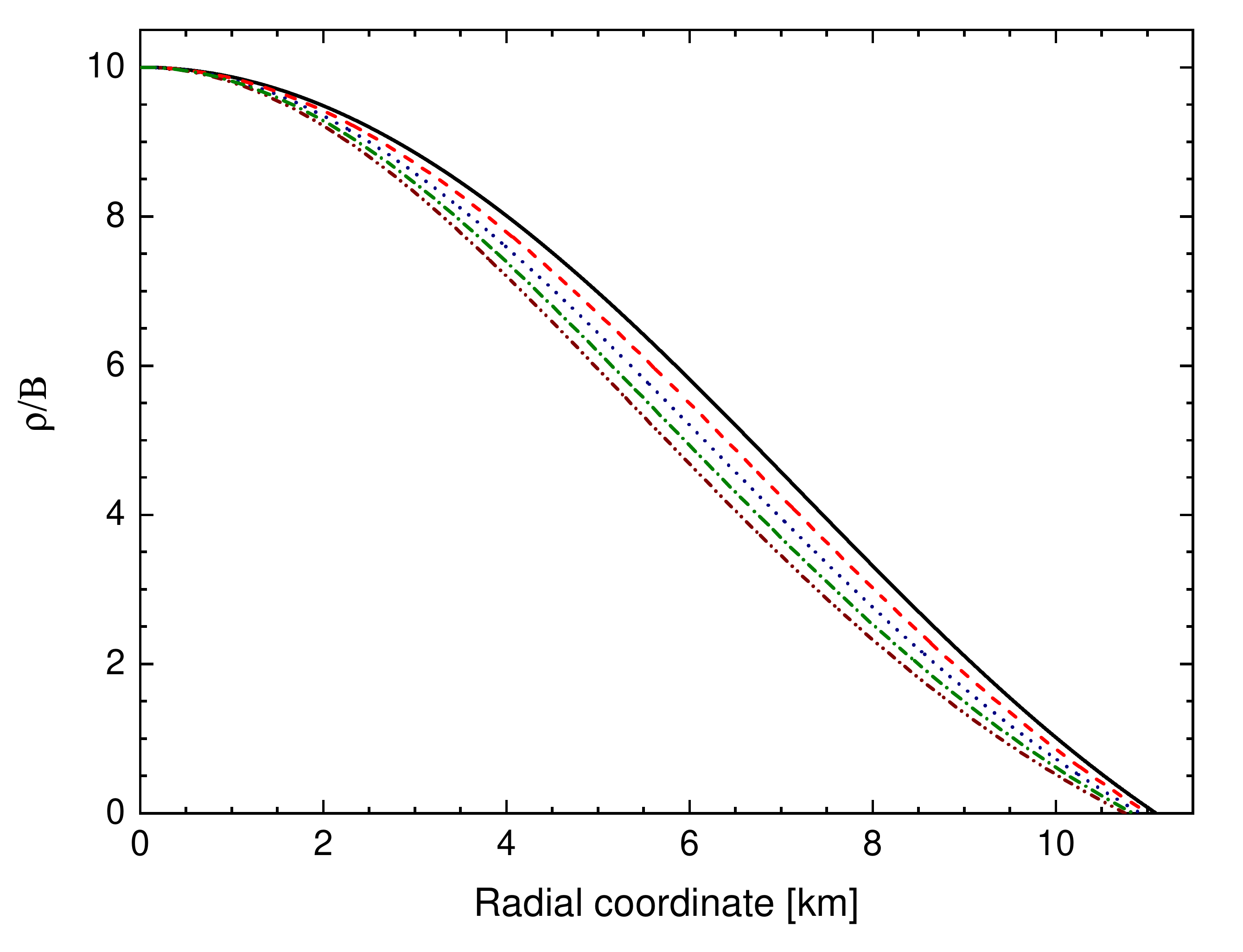} 
\includegraphics[width=8.5cm]{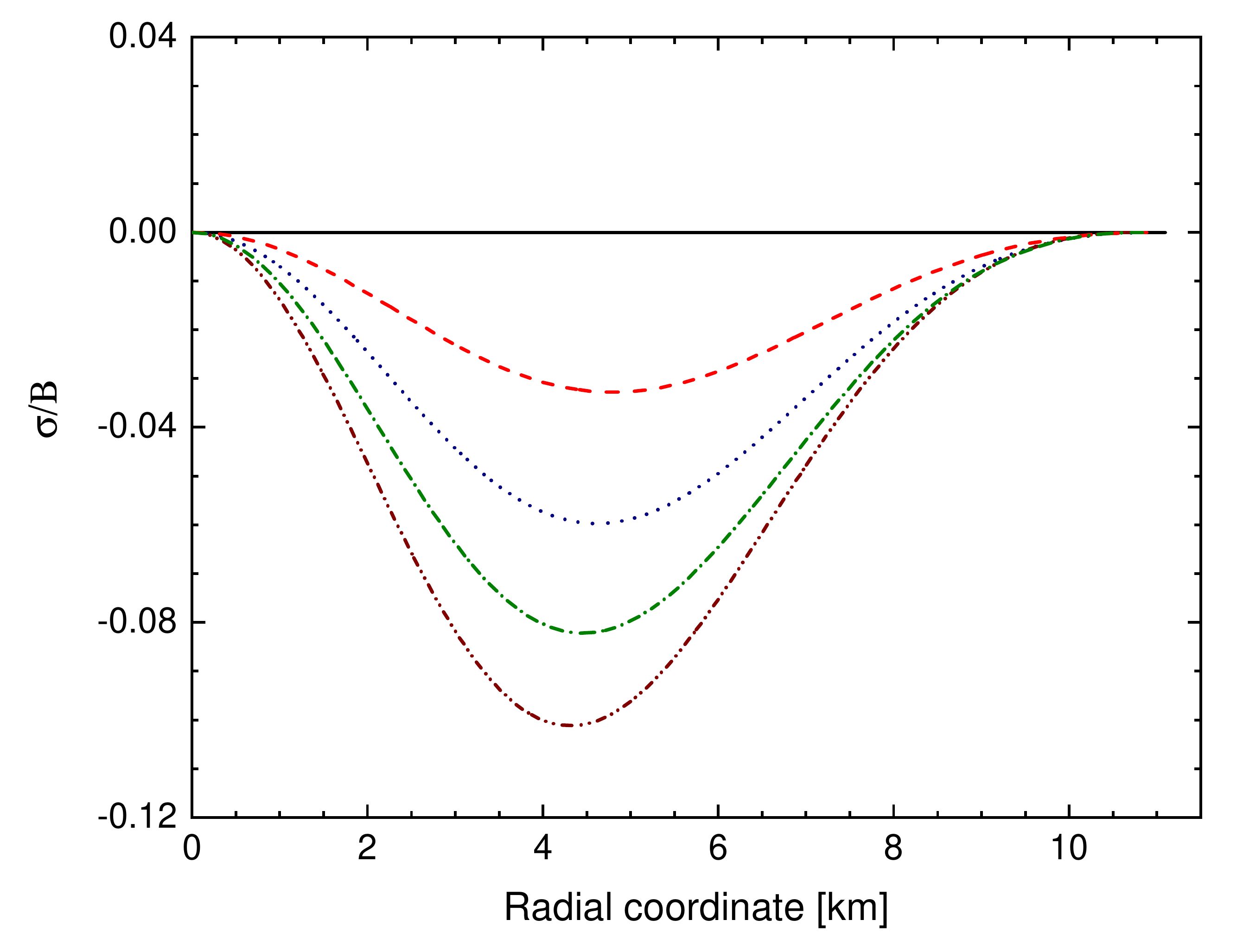} 
\includegraphics[width=8.5cm]{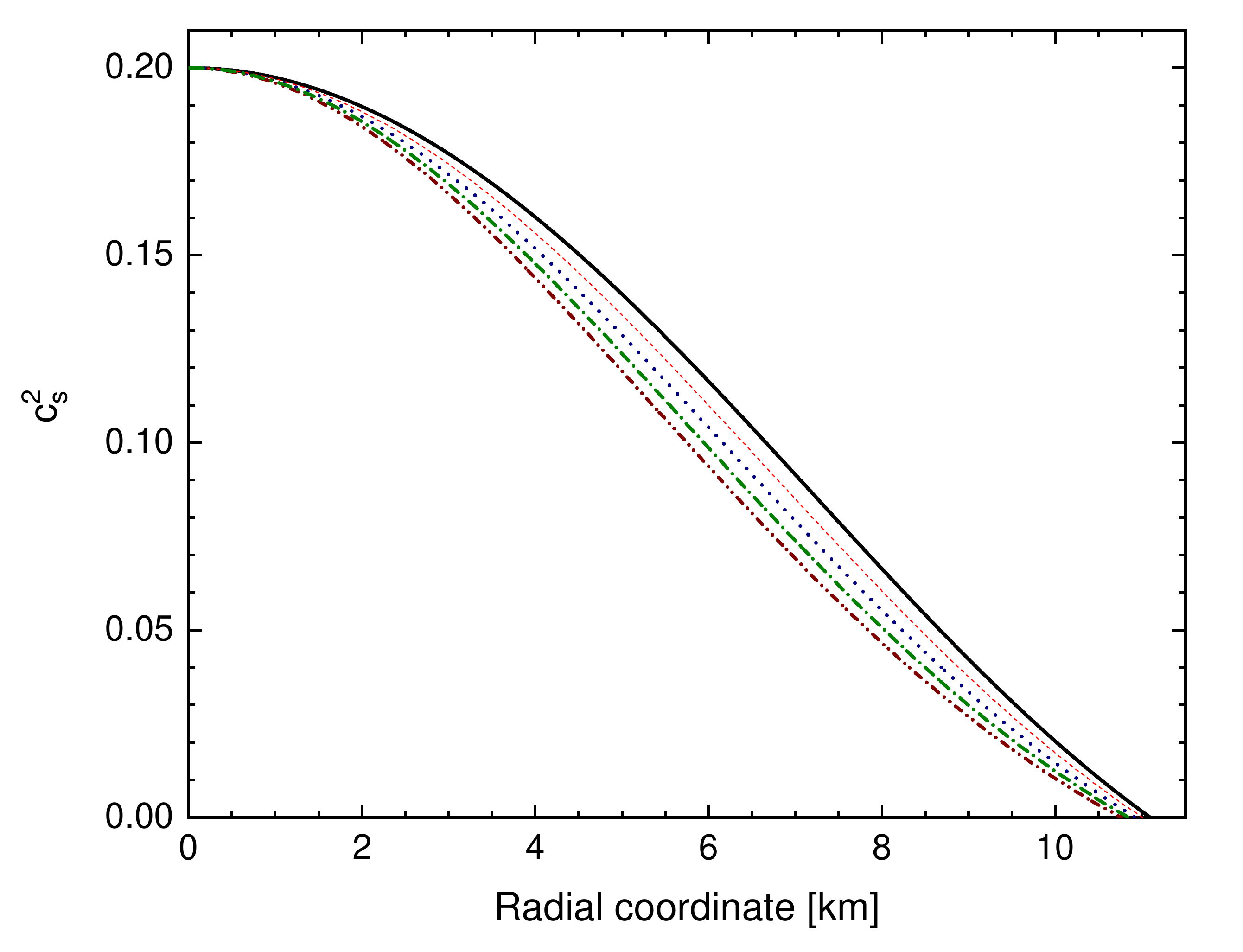} 
\caption{\label{fig:epsart1} Fluid radial pressure, energy density, anisotropic factor in their normalized form and the square sound velocity as a function of the radial coordinate for some different values of $\kappa$. $\rho_c=660\,[\rm MeV/fm^3]$ and $p_c=66\,[\rm MeV/fm^3]$ are considered in panels. The normalization factor is $B=66\,[\rm MeV/fm^3]$.}
\end{figure*}

\subsection{Numerical method}

The equilibrium configuration and tidal deformability of anisotropic polytropic spheres are analyzed by mean of the numerical integration of the system of equation and boundary conditions established in Section \ref{section2}. For each value of $\rho_c$ and $\kappa$ considered, this process starts in the center and end at the star's surface.

\smallskip

The study of the radial stability of anisotropic spheres begins solving the set of equations (\ref{eq_mass})-(\ref{eq_g00}), by employing the Runge-Kutta fourth-order method. Once determining the coefficients of the radial stability equations for each $\rho_c$ and $\kappa$, Eqs. \eqref{eq_ro_1} and \eqref{eq_ro_2} are solved through the shooting method for a trial value of $\omega^2$. If after each numerical integration the condition \eqref{delta_p_s} is not satisfied, the eigenfrequency squared is corrected with the aim to attain this equality in the next integration (check, e.g., \cite{arbanil_malheiro_2016}).


\subsection{Equilibrium of anisotropic polytropic spheres}


Fig.~\ref{fig:epsart1} shows the behavior of the fluid radial pressure $p_r/B$, energy density $\rho/B$, anisotropic factor $\sigma/B$ and the speed of sound squared $c_s^2$ with the radial coordinate for five values of $\kappa$. The central energy density and central pressure considered are respectively $\rho_c=660\,[\rm MeV/fm^3]$ and $p_c=66\,[\rm MeV/fm^3]$. The influence of the anisotropy in the radial pressure, energy density, and the speed of sound squared can be seen on panels. It is found the diminution of $\rho/B$, $p_r/B$, and $c_s^2$ with the diminution of $\kappa$.

\begin{figure}[h]
\centering
\includegraphics[width=8.5cm]{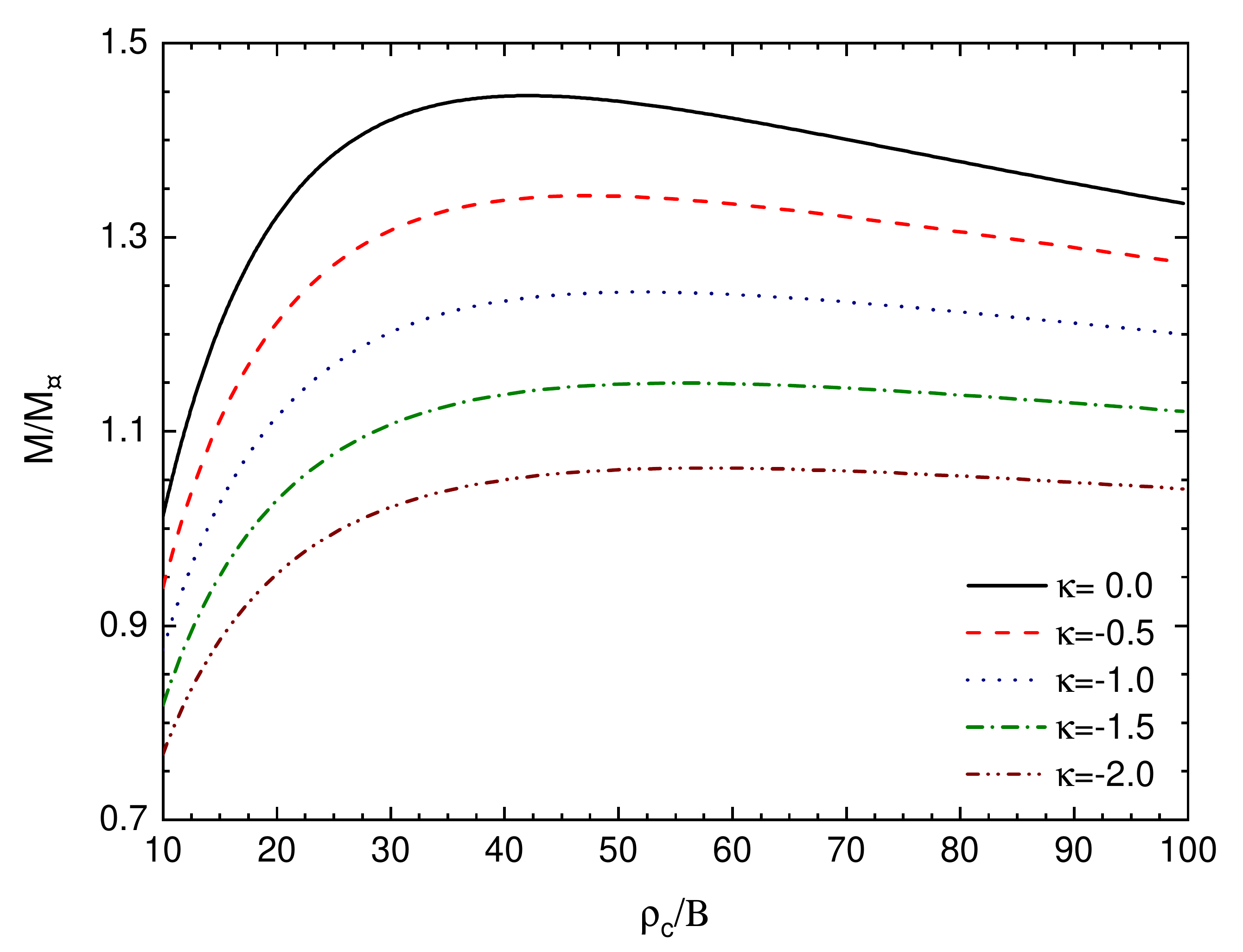} 
\caption{\label{fig:epsart2} The total mass, in Solar masses $M_{\odot}$, against the the normalized central energy density for few values of $\kappa$.}
\end{figure}

The mass, normalized to the Sun’s mass $M_{\odot}$, as a function of the central energy density is presented in Fig. \ref{fig:epsart2} for some values of $\kappa$. The central energy density considered are in the range $10\leq\rho_c/B\leq100$. In each curve, we find that the maximum mass point $M_{\rm max}/M_{\odot}$ and the null eigenfrequency of oscillation $\omega=0$ are determined by considering the same central energy density. From this result, we understand that the stable and unstable equilibrium configurations against small radial perturbation can be recognized by the conditions $dM/d\rho_c> 0$ and $dM/d\rho_c<0$, respectively. Some similar results are reported in literature. For instance, in \cite{hillebrandt_steinmetz_1976} and \cite{arbanil_malheiro_2016} are respectively investigated the stability of anisotropic neutron stars against radial perturbations, considering $\sigma=\kappa p_r$, and radial stability of anisotropic strange stars, employing $\sigma=\kappa p_r\left(1-e^{-\lambda}\right)$.


\begin{figure}[h]
\centering
\includegraphics[width=8.5cm]{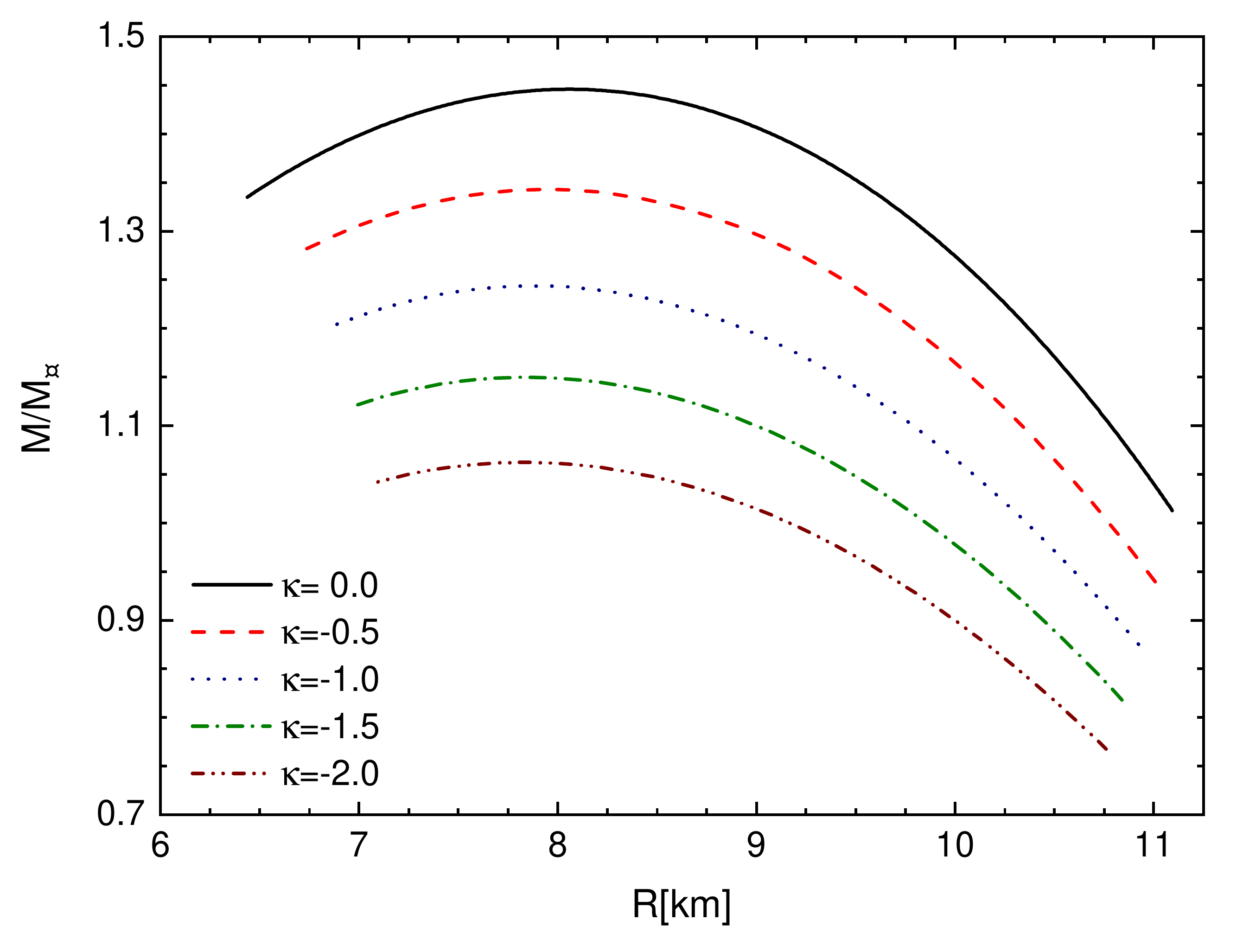} 
\caption{\label{fig:epsart3} The mass $M/M_{\odot}$ versus the radius for five values of $\kappa$.}
\end{figure}

The mass $M/M_{\odot}$ against the total radius $R$ for five different values of $\kappa$ is shown in Fig. \ref{fig:epsart3}. From curves, we observe that the mass and radius change substantially with $\kappa$. Once $\kappa<0$, for a lower $\kappa$ the compact object have smaller mass and total radius. The diminution of the mass with $\kappa$ can be understood inspecting the relation between $\sigma$ and $\kappa$. A lower negative anisotropy $\sigma$ counteracts the fluid pressure, thus obtaining equilibrium configurations with lower masses and radii.


\subsection{Radial stability of anisotropic polytropic spheres}

\begin{figure}[h]
\centering
\includegraphics[width=8.5cm]{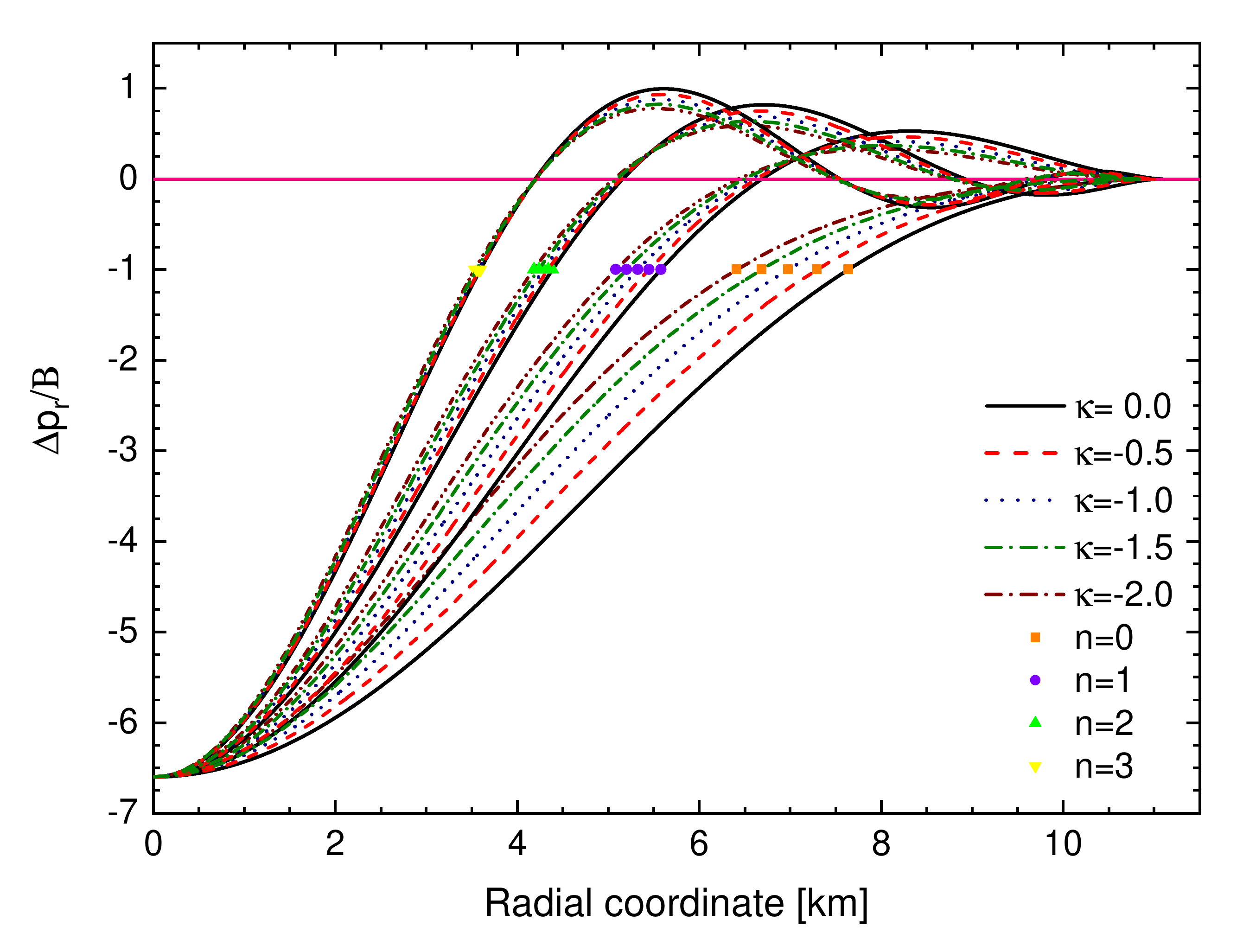} 
\includegraphics[width=8.5cm]{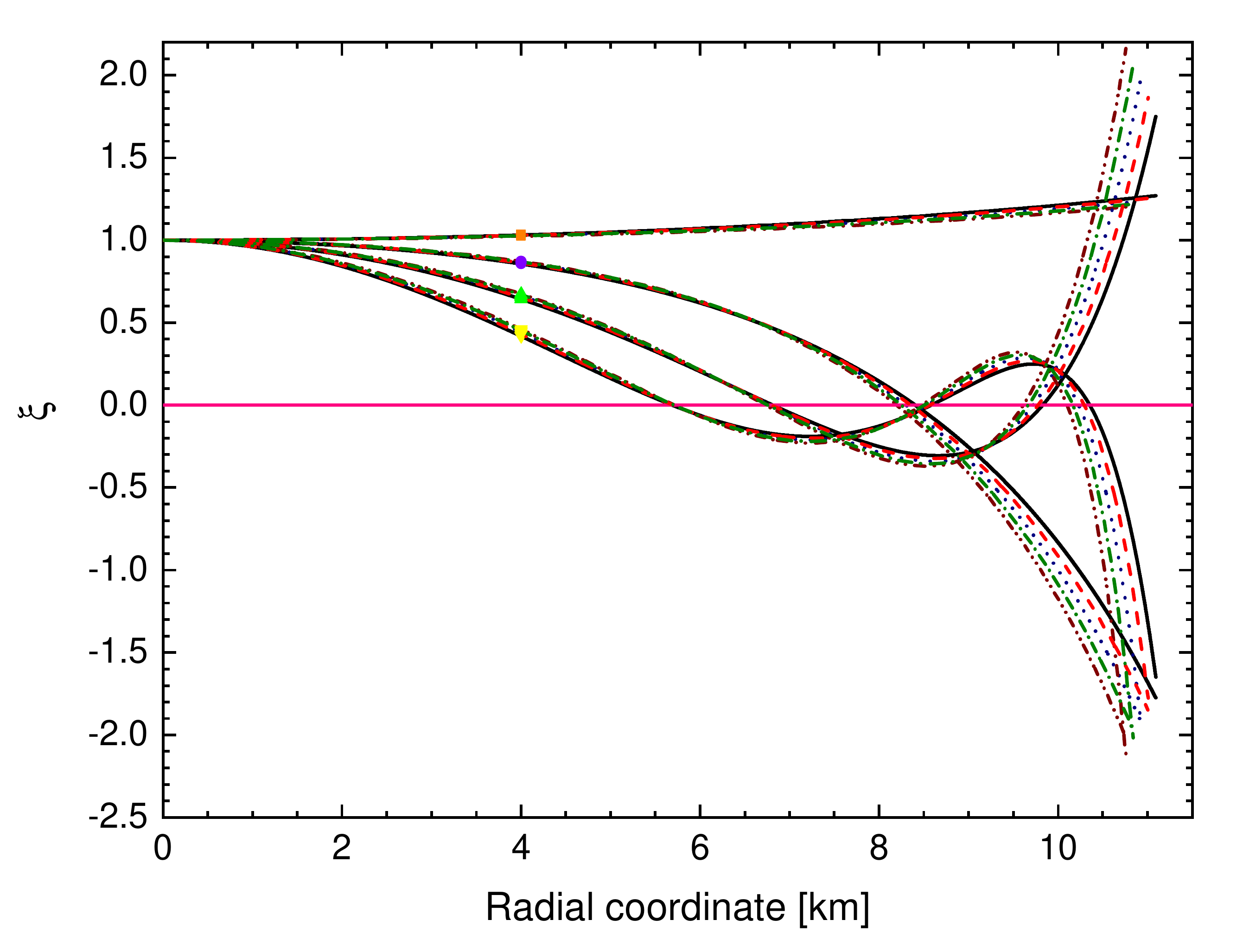} 
\caption{\label{fig:epsart4} The normalized Lagrangian perturbation $\Delta p_r/B$ and the relative radial displacement $\xi$ against the radial coordinate for some values of $\kappa$ and radial oscillation modes $n$.}
\end{figure}

The Lagrangian perturbation $\Delta p_r$ and the relative radial displacement $\xi$ against the radial coordinate $r$ are plotted on the top and bottom of Fig. \ref{fig:epsart4} for different $\kappa$ and overtones $n$. As can be seen in figure, before attain the star's surface, the fundamental mode ($n=0$) does not present any zeros while the first ($n=1$), the second ($n=2$) and third ($n=3$) exited mode present respectively one, two and three zeros.

\smallskip

The effects of anisotropy at the Lagrangian perturbation and relative radial displacement can also be seen in Fig. \ref{fig:epsart4}. We note that the effect of anisotropy is noticeable in each radial oscillation mode $n$. From these results, it can be understand that from radial pulsations of compact objects is possible to get some information about the internal structure and obtain some proof of the presence of anisotropy within compact objects.


\begin{figure}[h]
\centering
\includegraphics[width=8.5cm]{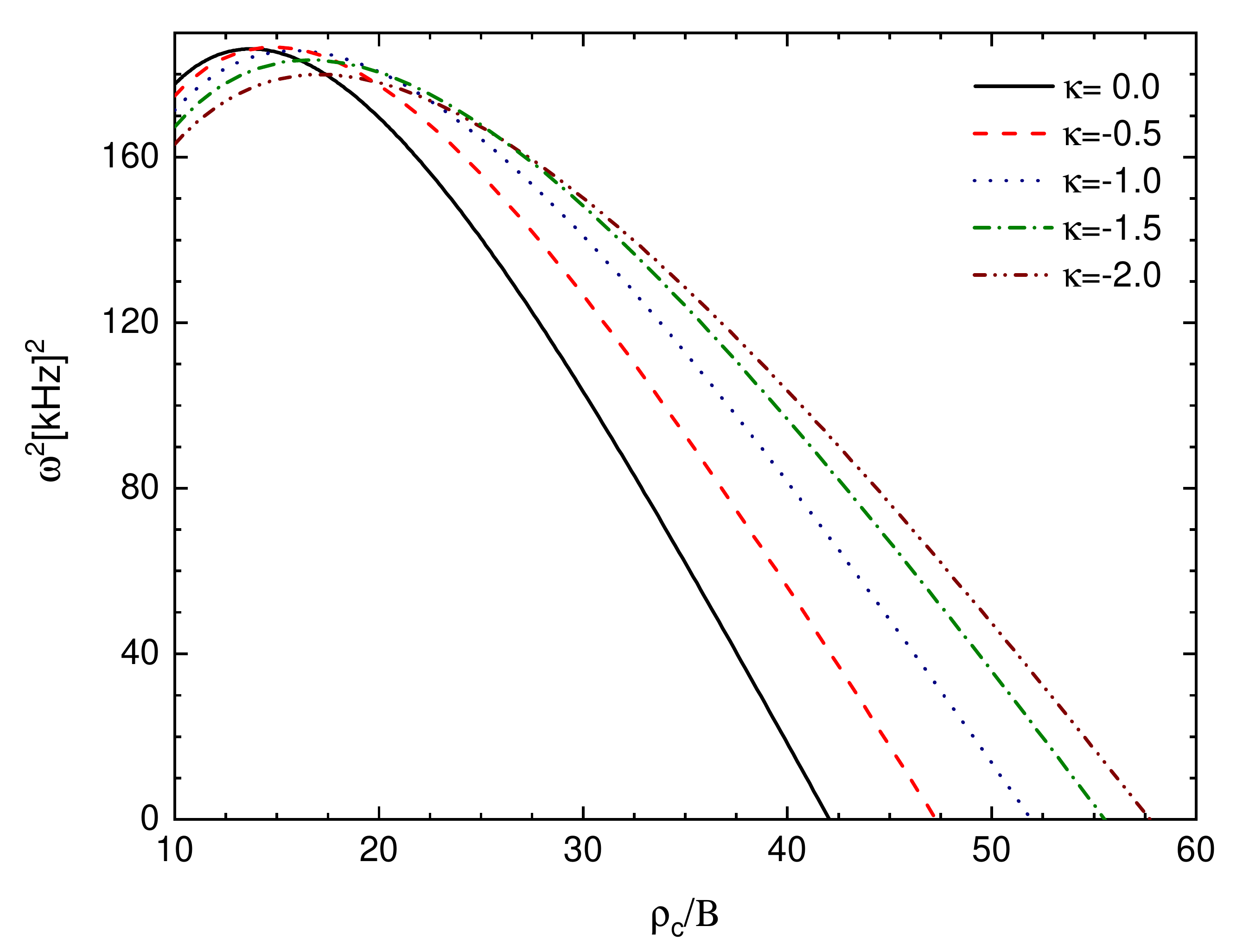} 
\caption{\label{fig:epsart5} The eigenfrequency of oscillation squared against the normalized central energy density for some values of $\kappa$.}
\end{figure}

The eigenfrequency of oscillation squared versus the central energy density normalized with $B$ is plotted in Fig. \ref{fig:epsart5} for few values of $\kappa$. Only radial stable equilibrium configurations are presented. In all cases shown, at small central energy densities, the curves $\omega^2(\rho_c)$ presents a small growth until attain a maximum eigenfrequency value, after this point $\omega^2$ decreases monotonically with the increment of $\rho_c$ thus displaying that a polytropic star with greater central energy density will have lesser radial stability. It can be also observed the effects of the anisotropy on the radial stability of the object. For an interval of central energy density, the diminution of $\kappa$ increases the radial stability of the spherical object.


\begin{figure}[h]
\centering
\includegraphics[width=8.5cm]{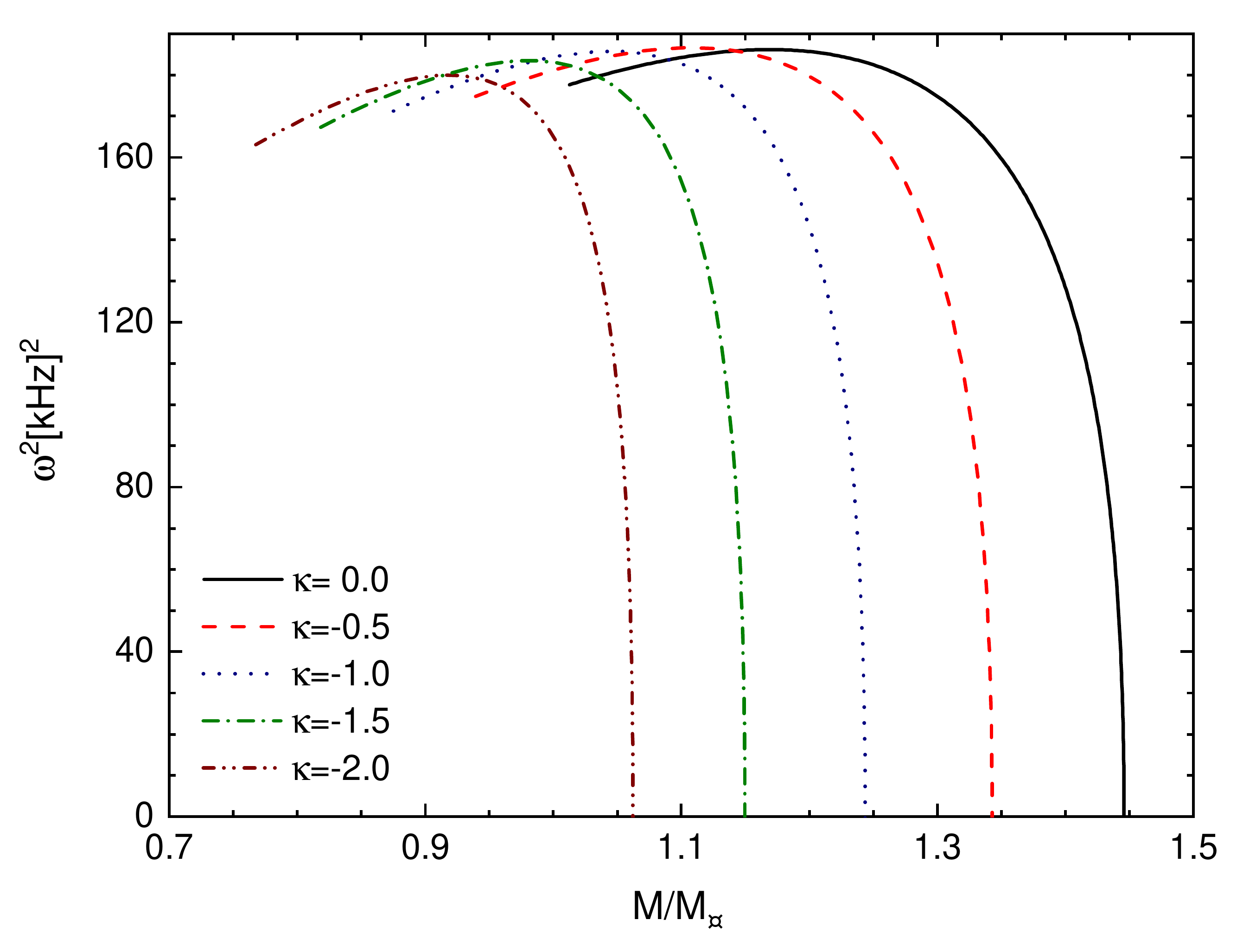} 
\caption{\label{fig:epsart6} The eigenfrequency of oscillation squared versus the total mass for five different values of $\kappa$.}
\end{figure}


Fig. \ref{fig:epsart6} shows the behavior of the eigenfrequency of oscillation squared with the total mass for some values of $\kappa$. In all cases, the curves $\omega^2(M)$ displays a little growth until reach a maximum eigenfrequency value, hereafter, $\omega^2$ decreases with the increment of the total mass until to find $\omega=0$ in $M_{\rm max}/M_{\odot}$. This maximum mass point marks the beginning of the radial instability. In addition, for some interval of total mass, the reduction of $\kappa$ induce a diminution of the radial stability.


\subsection{Tidal deformability of anisotropic polytropic spheres}

\begin{figure}[h]
\centering
\includegraphics[width=8.5cm]{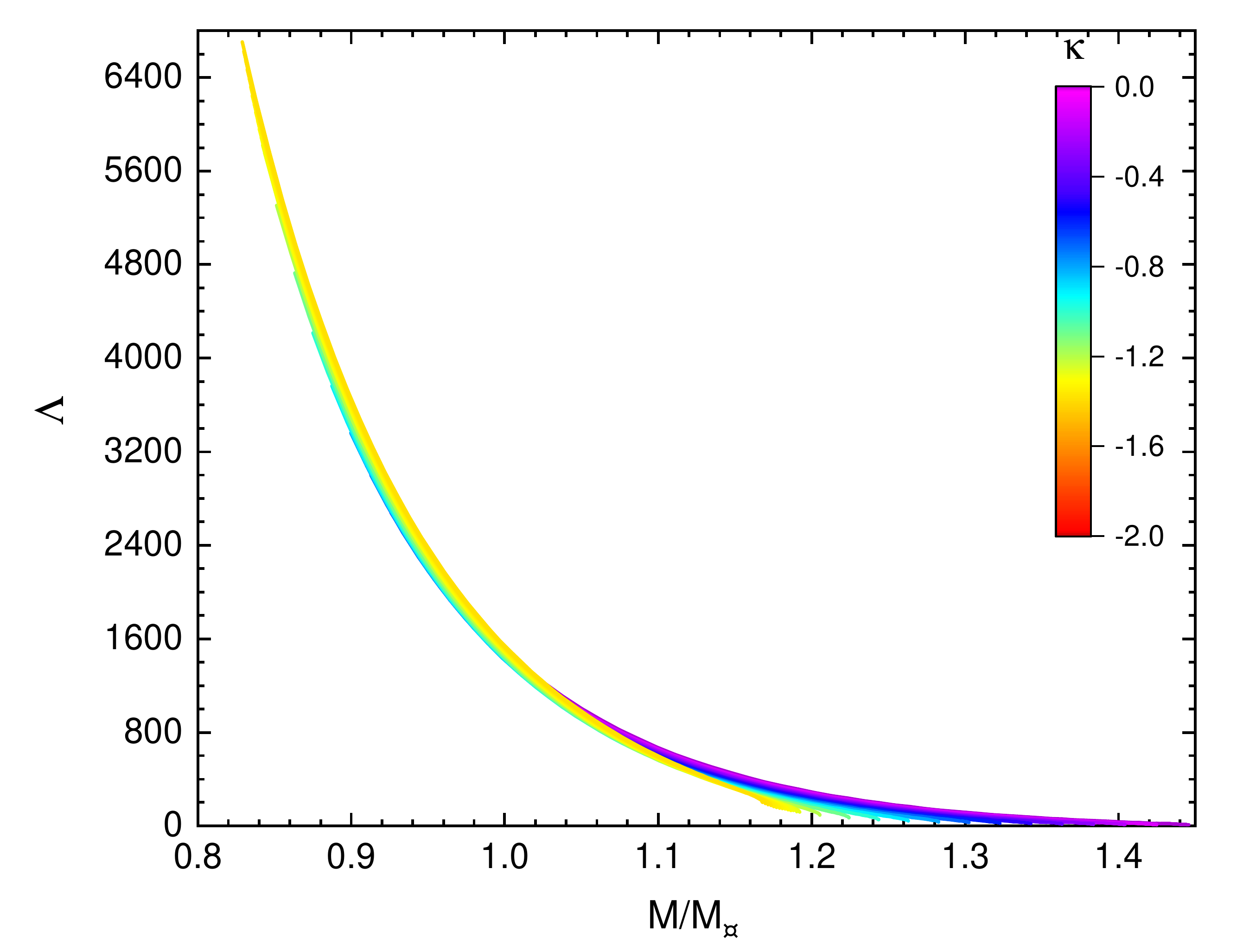} 
\includegraphics[width=8.5cm]{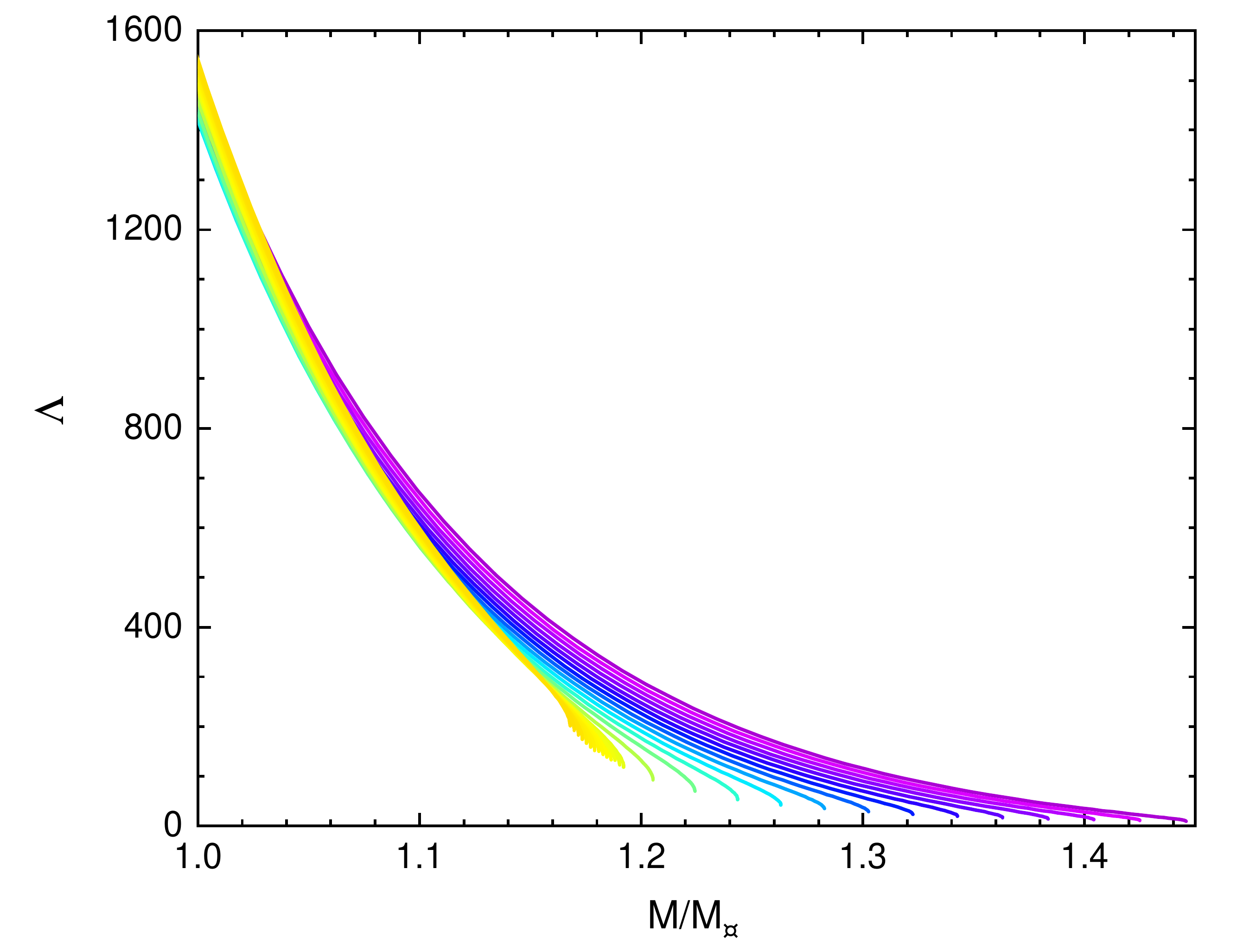} 
\caption{\label{fig:epsart7} Top: Dimensionless tidal deformability $\Lambda$ against the total mass in Solar masses. Bottom: Amplification of the region where the lowest values of tidal deformability are determined. Only stable equilibrium configurations are presented.}
\end{figure}

Tidal deformability $\Lambda$ as a function of the total mass is presented in Fig. \ref{fig:epsart7} for different values of $\kappa$. As can be noted, the deformability decreases with the increment of $M/M_{\odot}$ until to reach the maximum total mass. Furthermore, for an interval of total mass, $\Lambda$ decreases with $\kappa$. From this result, we can understand that the anisotropy in a compact object would significantly affect the tidal deformability. Finally, it is worth mentioning that the deformability of the last stable star increases with decreasing of the dimensionless prefactor $\kappa$.


\section{Conclusions}

The equilibrium configuration, the radial stability, and the tidal deformability of polytropic stars in the presence of anisotropy are investigated. This is realized by means of the numerical solution TOV equations, Chandrasekhar pulsation equations, and the Riccati equations for tidal deformability, all modified for the anisotropic case. For the matter that makes up the sphere, it follows the relation $p_r=K\rho^{2}$. On the other hand, the anisotropic factor employed takes the form $\sigma=\kappa p_r(1-e^{-\lambda})$.

\smallskip

We show how some properties of the compact star as the radial pressure, energy density, speed of sound squared, total mass, eigenfrequency of oscillation, and tidal deformability are affected by the anisotropy. 

\smallskip

Given the results, we note that for a central energy density interval the radial stability increases with the decrement of $\kappa$. Moreover, for a total mass range, the radial stability and tidal deformability decrease with $\kappa$. In all cases, we found that the maximum mass values and the zero eigenfrequencies of oscillations are derived by the same central energy densities, thus indicating that the maximum mass point indicates the beginning of radial instability. At this point, we obtained that the deformability of the last stable compact star grows with the diminution of the dimensionless prefactor.

Finally, it is important to mention that the change of the deformability with anisotropy could lead to the possibility that some equations of state that are outside of the limit of tidal deformability of the event GW$170817$, could be within this frame for certain values of the dimensionless prefactor (review also \cite{biswas2019}). Moreover, it can be noted that both the mass and the tidal deformability found for some EoSs, could help us to place restrictions on the value of $\kappa$.


\begin{acknowledgments}
\noindent JDVA thanks Universidad Privada del Norte and Universidad Nacional Mayor de San Marcos for the financial support - RR Nº$\,005753$-$2021$-R$/$UNMSM under the project number B$21131781$. The author G.P. thanks the Fun\-da\c c\~ao para a Ci\^encia e Tecnologia (FCT), Portugal, for the financial support to the Center for Astrophysics and Gravitation-CENTRA, Instituto Superior T\'ecnico, Universidade de Lisboa, through the Projects No.~UIDB/00099/2020 and No.~PTDC/FIS-AST/28920/2017.
\end{acknowledgments}


\end{document}